SEÇÃO: VARIA

# Modelos dinâmicos aplicados à aprendizagem de valores em inteligência artificial

*Dynamic models applied to learning values in artificial intelligence*
*Modelos dinámicos aplicados al aprendizaje de valores en inteligencia artificial*

**Nicholas Kluge Corrêa[1]**
orcid.org/0000-0002-5633-6094
nicholas.correa@acad.pucrs.br

**Nythamar de Oliveira[1]**
orcid.org/0000-0001-9241-1031
nythamar@yahoo.com

**Resumo:** especialistas em desenvolvimento de Inteligência Artificial (IA) preveem que o avanço no desenvolvimento de sistemas e agentes inteligentes irá remodelar áreas vitais em nossa sociedade. Contudo, se tal avanço não for realizado de maneira prudente e crítico-reflexiva, pode resultar em desfechos negativos para a humanidade. Por esse motivo, diversos pesquisadores na área têm desenvolvido uma concepção de IA robusta, benéfica e segura para a preservação da humanidade e do meio-ambiente. Atualmente, diversos dos problemas em aberto no campo de pesquisa em IA advêm da dificuldade de evitar comportamentos indesejados de agentes e sistemas inteligentes e, ao mesmo tempo, especificar o que realmente queremos que tais sistemas façam, especialmente quando prospectamos a possibilidade de agentes inteligentes atuarem em vários domínios em longo prazo. É de suma importância que agentes inteligentes artificiais tenham os seus valores alinhados com os valores humanos, dado ao fato de que não podemos esperar que uma IA desenvolva valores morais humanos por conta de sua inteligência, conforme é discutido na Tese da Ortogonalidade. Talvez tal dificuldade venha da maneira que estamos abordando o problema de expressar objetivos, valores e metas, utilizando métodos cognitivos representacionais. Uma solução para esse problema seria a abordagem dinâmica proposta por Dreyfus, que com base na filosofia fenomenológica mostra que a experiência humana do ser-no-mundo em diversos aspectos não é bem representada pelo método cognitivo simbólico ou conexionista, especialmente na questão de aprendizagem de valores. Uma possível abordagem para esse problema seria a utilização de modelos teóricos como *Situated Embodied Dynamics* (SED) para abordar o problema de aprendizagem de valores em IA.

**Palavras-chave:** Inteligência Artificial. Aprendizagem de valores. Ciência cognitiva. Sistemas Dinâmicos.

**Abstract:** experts in Artificial Intelligence (AI) development predict that advances in the development of intelligent systems and agents will reshape vital areas in our society. Nevertheless, if such an advance is not made prudently and critically-reflexively, it can result in negative outcomes for humanity. For this reason, several researchers in the area have developed a robust, beneficial, and safe concept of AI for the preservation of humanity and the environment. Currently, several of the open problems in the field of AI research arise from the difficulty of avoiding unwanted behaviors of intelligent agents and systems, and at the same time specifying what we really want such systems to do, especially when we look for the possibility of intelligent agents acting in several domains over the long term. It is of utmost importance that artificial intelligent agents have their values aligned with human values, given the fact that we cannot expect an AI to develop human moral values simply because of its intelligence, as discussed in the Orthogonality Thesis. Perhaps this difficulty comes from the way we are addressing the problem of expressing objectives, values, and ends, using representational cognitive methods. A solution to this problem would be the dynamic approach proposed by Dreyfus, whose phenomenological philosophy shows that the human experience of being-in-the-world in several aspects is not well represented by the symbolic or connectionist cognitive method, especially in regards to the question of learning values. A possible approach to this problem would be to use theoretical models such as SED (situated embodied dynamics) to address the values learning problem in AI.

---

[1] Pontifícia Universidade Católica do Rio Grande do Sul (PUCRS), Porto Alegre, RS, Brasil.

**Keywords:** Artificial Intelligence. Learning values. Cognitive science. Dynamical systems.

**Resumen:** los expertos en desarrollo de Inteligencia Artificial (IA) predicen que el progreso en el desarrollo de sistemas y agentes inteligentes remodelará áreas vitales en nuestra sociedad. Sin embargo, si ese progreso no se lleva a cabo de manera prudente y crítico-reflexiva, puede resultar en resultados negativos para la humanidad. Por esta razón, varios investigadores en el campo han desarrollado un diseño de IA robusta, beneficiosa y segura para la preservación de la humanidad y el medio ambiente. Actualmente, varios de los problemas abiertos en el campo de investigación de IA se derivan de la dificultad de evitar comportamientos no deseados de agentes y sistemas inteligentes, y al mismo tiempo especificar lo que realmente queremos que hagan estos sistemas, especialmente cuando prospectemos la posibilidad de que agentes inteligentes operen en varias áreas a largo plazo. Es de suma importancia que los agentes inteligentes artificiales tengan sus valores alineados con los valores humanos, dado el hecho de que no podemos esperar que una IA desarrolle valores morales humanos debido a su inteligencia, como se discutió en la Tesis de Ortogonalidad. Tal vez esta dificultad proviene de la forma en que estamos abordando el problema de expresar objetivos, valores y objetivos, utilizando métodos cognitivos representativos. Una solución a este problema sería el enfoque dinámico propuesto por Dreyfus, que basado en la filosofía fenomenológica muestra que la experiencia humana del ser-en-el-mundo en diversos aspectos no está bien representada por el método cognitivo simbólico o conexionista, especialmente en la cuestión del aprendizaje de valores. Un posible enfoque de este problema sería utilizar modelos teóricos como (SED) (situated embodied dynamics) para abordar el problema del los valores en la IA.

**Palabras clave:** Inteligencia Artificial. Aprendizaje de valores. Ciencia cognitiva. Sistemas dinámicos.

## Introdução

Pesquisadores e especialistas em desenvolvimento de Inteligência Artificial (IA) estipulam que dentro de 10 anos muitas atividades humanas serão superadas por máquinas em termos de eficiência. Diversos aspectos de nossas políticas públicas precisarão se modificar para acompanhar tais avanços, que prometem remodelar áreas como transporte, saúde, economia, combates militares, estilo de vida etc. (GRACE *et al.*, 2017). Existe, também, a preocupação acerca dos riscos que máquinas com alto nível de inteligência humana ou sobre-humana podem trazer para a humanidade nas próximas décadas. Uma pesquisa realizada por Müller e Bostrom (2016) consistiu em construir um questionário para avaliar o progresso do campo de pesquisa em IA e as prospecções para o futuro, entrevistando diversos especialistas na área. O questionário mostrou que, em média, há uma chance de 50% de que a inteligência de máquina de alto nível (humana) seja alcançada entre 2040 e 2050, chegando a 90% de probabilidade até 2075. Estima-se também que esta inteligência superará a performance humana em 2 anos (10% de chance) a 30 anos (75% de chance) após atingir níveis de inteligência humana (MÜLLER; BOSTROM, 2016).

Contudo, na mesma pesquisa, 33% dos entrevistados classificaram esse desenvolvimento em IA como "ruim" ou "extremamente ruim" para a humanidade (MÜLLER; BOSTROM, 2016). Como não há nenhuma garantia de que tais sistemas serão "bons" para a humanidade, devemos investigar melhor o futuro da superinteligência e os riscos que ela representa à raça humana. Existem, portanto, várias questões em aberto e problemas que necessitam de solução. Como remediaremos os impactos econômicos da IA com o intuito de evitar efeitos negativos como o desemprego em massa? (FREY; ORSBORNE, 2013). Como evitar que a automação dos empregos empurre a distribuição de renda para uma lei de poder desproporcional entre classes, gêneros e raças? (BRYNJOLFSSON; MCAFEE, 2014). Armas letais autônomas podem ser construídas sem modificar os direitos humanitários? (CHURCHILL; ULFSTEIN, 2000). Deveriam armas autônomas serem completamente banidas? (DOCHERTY, 2012). Como podemos garantir a privacidade ao aplicar o aprendizado de máquina a dados confidenciais como fontes de dados médicos, linhas telefônicas, e-mails, padrões de comportamento on-line (ABADI *et al.*, 2016)? Como podemos entender o que os sistemas de IA complexos estão fazendo para classificar e reconstruir de forma iterativa imagens a partir de redes neurais? (MORDVINTSEV; OLAH; TYKA, 2015).

Alguns pesquisadores já criaram modelos (ASI-PATH) de como uma IA poderia vir a causar algum tipo de catástrofe, tornando-se superinteligente através do autoaperfeiçoamento recursivo (BARRET; BAUM, 2017), algo conhecido na literatura de IA como o Paradoxo da Singularidade. Tais modelos sugerem cenários onde agentes inteligentes, após obterem algum tipo de vantagem estratégica, (*decisive strategic advantage* [DAS] ou *major strategic advantage* [MAS]) como, por exemplo, avanços em nanotecnologia ou robótica, poderiam alcançar considerável poder de dominação (BOSTROM; ĆIRKOVIĆ, 2008). Os cenários sugerem tipos diferentes de tomadas de controle por sistemas inteligentes artificiais, variando de tomadas rápidas (*fast takeoff*), situações em que ocorre uma drástica tomada de poder por tais sistemas, a tomadas lentas (*slow takeoff*), onde gradualmente a raça humana se torna mais dependente e, de certa forma, sob o controle de IA (SOTALA, 2018). O desenvolvimento de uma ética da IA pressupõe, com efeito, as formulações intuitivas das chamadas Três Leis da Robótica de Isaac Asimov (1950), em uma época em que esse tema ainda parecia relegado ao domínio da *science fiction* –lembrando que tais codificações ético-morais foram introduzidas em um conto de 1942, *Runaround*: "(1) Um robô não pode ferir um ser humano ou, por inação, permitir que um ser humano seja prejudicado; (2) um robô deve obedecer às ordens dadas pelos seres humanos, exceto onde essas ordens entrem em conflito com a Primeira Lei; (3) um robô deve proteger sua própria existência, desde que essa proteção não entre em conflito com a Primeira ou a Segunda Lei." Em nosso século, essa orientação ética de não causar dano à humanidade foi estendida não apenas a robôs e artefatos robóticos, mas a máquinas e dispositivos inteligentes de forma genérica associados a recursos de IA. Assim, Shulman (2010, p. 2) sugere um modelo que explica em quais situações uma IA iria abandonar a cooperação com a raça humana e tomar medidas hostis, no qual um agente artificial que acredita ter probabilidade P de ser bem sucedido, se iniciar agressão, recebendo utilidade [EU (Sucesso)], e com probabilidade (1 - P) de falhar como essa ação, recebendo [EU (Fracasso)]. Caso desista de uma estratégia agressiva, o agente recebe utilidade [EU (Cooperação)]. A IA irá iniciar de forma racional a agressão apenas se: P×EU(Sucesso)+(1-P)×EU(Fracasso)> EU(Cooperação).

## Problemas de segurança em IA

Nota-se, em última análise, que existe um consenso na literatura: é extremamente importante que o desenvolvimento de IA seja feito de maneira segura, benéfica e robusta. Um artigo publicado por Amodei *et al.* (2016), intitulado "*Concrete Problems in AI Safety*", enumera uma série de problemas em aberto no campo de pesquisa em IA que devem ser solucionados para que possamos colher os benefícios da IA sem comprometer nossa segurança. Estes problemas são classificados em problemas de especificação ou robustez, e são as atuais barreiras a serem vencidas na área (LEIKE *et al.*, 2017).

Para melhor sintetizar e desenvolver o conteúdo, nós iremos nos referir de maneira breve apenas aos erros de especificação. Erros de especificação ocorrem quando a função de utilidade da IA é mal especificada pelos programadores, causando resultados indesejados e até mesmo prejudiciais, mesmo que o aprendizado seja perfeito com dados explicitamente claros (AMODEI *et al.*, 2016). Alguns exemplos de erros de especificação são efeitos colaterais negativos, hackeamento de recompensa e interrupção segura (corrigibilidade).

Efeitos colaterais negativos acontecem quando a maximização da função de recompensa se concentra em realizar um objetivo enquanto o agente ignora fatores importantes no ambiente, causando potenciais efeitos transversais. Em hackeamento de recompensa, o agente de IA encontra uma solução para seu objetivo que maximiza sua função de recompensa, mas de maneira inesperada, pervertendo a intenção dos programadores (AMODEI *et al.*, 2016). Já a Interrupção Segura ou Corrigibilidade diz respeito a como podemos ser capazes de interromper um agente caso ele esteja se comportando de maneira inesperada, e de certa forma, corrigir os erros detectados sem que o agente se oponha a interrupções (SOARES *et al.*, 2015).

Duas teses publicadas por Bostrom, (2012), apontam como esses problemas podem ser temerários. A "Tese da Convergência Intrumental" nos mostra como uma série de objetivos de autoaperfeiçoamento e preservação podem ser perseguidos por qualquer agente inteligente com um objetivo terminal. Podemos formular esta tese da seguinte maneira:

> Vários objetivos instrumentais podem ser identificados, os quais são convergentes no sentido de que sua obtenção aumentaria as chances do objetivo terminal do agente, implicando que esses objetivos instrumentais provavelmente serão perseguidos por qualquer agente inteligente (BOSTROM, 2012, p. 6).

Sem uma engenharia cuidadosa desses sistemas, riscos com uma "explosão de inteligência" (o aumento exponencial da capacidade cognitiva do agente) podem criar agentes muito mais poderosos que a nossa capacidade de os controlar. Por outro lado, e correlata à primeira tese, a "Tese da Ortogonalidade" propõe que a inteligência e os objetivos finais possuem propriedades independentes e ortogonais. A hipótese é argumentada da seguinte forma:

> Inteligência e objetivos finais são eixos ortogonais ao longo dos quais possíveis agentes podem variar livremente. Em outras palavras, mais ou menos qualquer nível de inteligência poderia, em princípio, ser combinado com mais ou menos qualquer objetivo final (BOSTROM, 2012, p. 3).

O pensamento por trás da tese da ortogonalidade é análogo à chamada Guilhotina de Hume (também conhecida, em inglês, como *Hume's fork* ou *Hume's law*), contrapondo o que é factual e empiricamente constatável (*matters of fact and real existence*) ao que deve ser, em termos racionais, normativos e contrafactuais (*relations of ideas*). Hume observou uma diferença significativa entre afirmações descritivas e afirmações prescritivas ou normativas, e que, portanto, não seria óbvio, autoevidente (*self-evident*) ou válido (*valid*) derivar as últimas das primeiras. A passagem indevida do ser (*Is*) ao dever-ser (*Ought*), que seria um dos problemas seminais das pesquisas em metaética, ética normativa e ética aplicada no século XX, foi constatada pelo filósofo escocês em uma famosa passagem da seção I da parte I do livro III de seu *Treatise of Human Nature*:

> Em todo sistema de moral que até hoje encontrei, sempre notei que o autor segue durante algum tempo o modo comum de raciocinar, estabelecendo a existência de Deus, ou fazendo observações a respeito dos assuntos humanos, quando, de repente, surpreendo-me ao ver que, em vez das cópulas proposicionais usuais, como é e não é, não encontro uma só proposição que não esteja conectada a outra por um deve ou não deve. Essa mudança é imperceptível, porém da maior importância. Pois como esse deve ou não deve expressa uma nova relação ou afirmação, esta precisaria ser notada e explicada; ao mesmo tempo, seria preciso que se desse uma razão para algo que parece totalmente inconcebível, ou seja, como essa nova relação pode ser deduzida de outras inteiramente diferentes (HUME, 2000, p. 509).

Assim como enunciados descritivos, puramente factuais, podem apenas vincular ou implicar outros enunciados descritivos ou factuais e nunca normas, pronunciamentos éticos ou prescrições para fazer alguma coisa, podemos também tentar inutilmente articular problemas de ortogonalidade e de alinhamento de valores, onde o problema de alinhamento de valores consistiria em garantir, caso uma AGI (*artificial general intelligence*, isto é, uma inteligência hipotética de uma máquina com capacidade de entender ou aprender qualquer tarefa intelectual que um ser humano possa realizar) chegasse a desenvolver inteligência suficiente para ter poder sobre a espécie humana, que tal inteligência faça apenas com os seres humanos o que desejaríamos ou aceitaríamos que fosse feito. Nesse sentido, o problema do alinhamento é idêntico ao que vemos na filosofia moral com relação ao utilitarismo, na medida em que a maximização da utilidade por parte de algum agente moral pode culminar em conclusões moralmente repugnantes, incluindo a violação dos direitos dos outros. Embora possa garantir a resolução de tarefas em tempo computacional (polinomial), a mera eficiência ou otimização de procedimentos não assegura a universalizabilidade normativa (como seria, de resto, uma premissa básica de modelos éticos deontológicos e não utilitaristas) e pode eventualmente entrar em conflito com os interesses ou direitos de outras pessoas. Cabe-nos ainda observar que a ética da inteligência artificial faz a parte da ética da tecnologia em geral e, de forma específica, para robôs, máquinas de aprendizado (*learning machines*) e outros artefatos e entes artificialmente inteligentes. Em nossa abordagem, a ética da AI compreende tanto uma roboética (ética robótica), que se preocupa com o comportamento moral dos seres humanos ao projetar, construir, usar e programar seres artificialmente inteligentes e uma ética de máquinas, que se preocupa com o comportamento moral dos próprios agentes morais artificiais. Tanto a bioética quanto a neuroética teriam muito a aprender, a ensinar e a interagir com a ética da inteligência artificial, sobretudo pela interface de modelos de vida artificial, edição genômica e redes neurais com os desafios ético-normativos da ortogonalidade, do alinhamento de valores e do transumanismo, integrando os legados neurobiológicos, culturais e tecnológicos do *homo sapiens sapiens*.

Pronunciamentos éticos e prescrições para o que se deve fazer não podem ser alcançados através da análise factual, havendo assim uma certa independência entre a razão e a moralidade, análogo à tese da ortogonalidade contrapondo inteligência e valores. Dessa forma, mesmo tendo uma função de utilidade "perfeita" implantada em um agente superinteligente, não podemos prever seus meios de atuação (objetivos instrumentais) e muito menos esperar que tal IA possua valores alinhados com os nossos. Agentes extremamente inteligentes podem ter diversos tipos de objetivos terminais acompanhados com uma grande gama de objetivos instrumentais, portanto, não devemos ceder à tentação de antropomorfizar uma IA (BOSTROM, 2012).

O viés antropomórfico tende a moldar todo o espectro de mentes e inteligências possíveis, mas isto é um engano, conhecido como a Falácia da Projeção da Mente (JAYNES, 2003). Pelo contrário, consideramos a inteligência como uma função de otimização da capacidade de um agente para atingir metas em uma ampla variedade de ambientes com recursos limitados (LEGG, 2008). Para melhor exemplificar este pensamento, utilizamos uma citação de Dijkstra, (1984), "a questão de saber se uma máquina pode pensar é tão relevante quanto a questão de saber se os submarinos sabem nadar". No sentido kantiano, a razão pode ser definida como a capacidade de obter inferências lógicas ou, de modo sistemático, a faculdade de sintetizar na unidade, por meio de princípios abrangentes, os conceitos fornecidos pelo intelecto, na medida em que apenas os seres humanos usam a razão para estabelecer e buscar fins (metas, propósitos, *Zwecken*), usando o resto da natureza como meio para seus fins. A humanidade é, assim, considerada como um fim em si e fim terminal da natureza, reconciliando os usos teórico-especulativo e prático da razão (ALLISON, 1996).

O limite superior de processamento bruto para todo o universo conhecido, imposto pelas leis da física, é de $10^{120}$ operações em $10^{90}$ bits ($10^{120}$ bits incluindo os graus de liberdade gravitacional) (LLOYD, 2002). O nível humano de processamento de informação é de $10^{11}$ operações por segundo (MORAVEC, 1998). Essa diferença entre o nível humano e o mais alto grau de otimização possível deixa em aberto uma vasta gama de possíveis níveis de inteligência sobre-humana (SOTALA, 2010).

Por estas razões, o alinhamento de valores entre IA e humanos é um importante problema a ser solucionado na área de ética da máquina (SOARES; FALLESTEIN, 2014). Praticamente todos os problemas de especificação, robustez e alinhamento de valores parecem ocorrer no mesmo ponto, quando nossas representações de valores ou objetivos finais (metas) perdem seu significado, ou são interpretadas erroneamente. Seria a abordagem objetiva-representacional fadada ao erro? Seriam os modelos cognitivos usados na criação de agentes inteligentes artificiais, especialmente simbolismo e conexionismo, incapazes de expressar o significado dos valores humanos? Se sim, haveria alguma alternativa?

## Modelos cognitivos: simbolismo e conexionismo

Desde o final da década de 1950, a discussão sobre cognição e inteligência tem sido permeada pela visão "computacional", também conhecida como visão simbólica. Esta perspectiva parte do pressuposto que sistemas cognitivos são inteligentes na medida em que podem codificar conhecimento em representações simbólicas. Simbolistas acreditam que através de conjuntos de regras de "se-então" (*if-then*) e outras formas de cálculo para algoritmos simbólicos, toda a cognição é realizada pela manipulação de tais representações (THAGARD, 1992).

Newell, (1990), definiu a proposta computacionalista, ao qual é referida também como a "Hipótese do Sistema de Símbolo Físico", da seguinte forma: Os sistemas cognitivos naturais são inteligentes em virtude de serem sistemas físicos que manipulam símbolos de forma a apresentar comportamento inteligente, codificando o conhecimento sobre o mundo externo em estruturas simbólicas (NEWELL, 1990, p. 75-79). Newell dedicou grande parte do seu trabalho na construção de sistemas que expressam sua visão de sistema de símbolo físico. Seu modelo mais promissor é conhecido como SOAR. SOAR é um sistema simbólico computacional que formula suas tarefas com base em hierarquias de símbolos e metas, gerando assim um sistema de produção e tomada de decisão algorítmica para resolução de problemas (NEWELL, 1990, p. 39).

Já no modelo conexionista propriedades emergentes, como o fenômeno de cognição, são efeitos de alto nível que dependem de fenômenos de nível inferior. Dessa forma, a hipótese conexionista encapsula a ideia de que o fato que mais determina a capacidade cognitiva de um agente não é a habilidade de manipulação representacional, mas sim sua arquitetura. Destarte, conexionistas atacam o problema da cognição realizando engenharia reversa no sistema nervoso central, copiando sua unidade básica de processamento, a saber, o neurônio (CHURCHLAND; SEJNOWSKI, 1992, p. 2). Sejnowski (1988, p. 7) nota em sua hipótese conexionista: "O processador intuitivo é um sistema dinâmico conexionista subconceitual que não admite uma completa, formal e precisa descrição em nível conceitual".

Assim, teorias de cognição em IA (simbolismo, conexionismo e dinamismo) podem ser consideradas estruturas teóricas, pois nos fornecem os filtros, analogias e metáforas pelas quais tentamos compreender o fenômeno de cognição, e assim criarmos modelos teóricos que possam gerar simulações a serem testadas (BEER, 1998). O simbolismo, por exemplo, ressalta as representações internas do sistema ou agente, e os algoritmos pelos quais essas representações são manipuladas. Já o conexionismo enfatiza a arquitetura da rede neural, o algoritmo de aprendizagem, a preparação dos dados de treinamento e o protocolo usado (ELIASMITH, 1996).

Todavia, as limitações da hipótese computacional simbolista, especialmente nos aspectos de tempo, arquitetura, computação e representação, levaram pesquisadores a cogitar novos modelos teóricos, como a hipótese dinâmica (VAN GELDER, 1998). E por mais que o modelo conexionista se assemelhe com o modelo dinâmico nos aspectos apontados no modelo simbólico (tempo, arquitetura, computação e representação), o conexionismo ainda falha em produzir agentes que consigam sanar os problemas de especificação e robustez citados anteriormente.

Neste artigo, não adotamos uma posição antirrepresentacionalista, na medida em que nós humanos utilizamos e manipulamos representações constantemente, como na linguagem, escrita, fala, música e outras formas de pensamento abstrato. Contudo, nos posicionamos ceticamente em relação à função de representações em sistemas que envolvam valores-objetivos-metas, e, portanto, comportamento guiado por metas. Talvez, em alguns casos, os papéis desempenhados pelos estados internos de um agente cognitivo simplesmente não podem ser interpretados como representacionais (FRANKISH; RAMSEY, 2014).

### Crítica ao Método Simbólico

Uma das maiores críticas levantadas contra o modelo computacional simbólico é a dificuldade em cumprir restrições temporais. Ao tentar replicar o fenômeno de cognição van Gelder

e Port, (1998, p. 2) afirma que os simbolistas "deixam o tempo fora de cena". Sendo o objetivo da ciência cognitiva descrever o comportamento de agentes cognitivos naturais, e por definição esses agentes operam em tempo real, um modelo cognitivo que replique a experiência humana de cognição deve apresentar processos cognitivos em tempo real (no caso de humanos: ± 10 milissegundos) (VAN GELDER; PORT, 1998).

Já os limites impostos pela arquitetura simbólica são uma outra fonte de crítica ao método computacional. Para Newell, (NEWELL, 1990, p. 82), o comportamento é determinado por um conteúdo variável sendo processado por uma estrutura fixa, que é a arquitetura. Dinamicistas criticam essa visão do sistema cognitivo como "uma caixa" dentro de um corpo, por sua vez dentro de um ambiente físico. Contudo, onde traçamos a linha que divide a "caixa" do seu corpo? E, mais controversamente, do corpo com o ambiente? Van Geldere e Port, (1998, p. 8), analisam pelo viés dinâmico a arquitetura interna no agente cognitivo como não sendo uma estrutura fixa, onde todos os aspectos da cognição, cérebro-corpo-ambiente, como mutuamente influenciando um ao outro continuamente.

Consequentemente, essa visão de arquitetura faz com que o método simbólico seja muitas vezes referido como método computacional, pois descreve a mente como um tipo especial de computador. Essa caracterização está totalmente de acordo com a arquitetura proposta por Newell, (1990), e identifica o computador mental com o cérebro. O corpo, através dos órgãos sensoriais, entrega até o sistema cognitivo (cérebro) representações do estado de seu ambiente; o sistema por sua parte calcula uma resposta apropriada. O corpo carrega essa ação (VAN GELDER; PORT, 1998, p. 1). Contudo, esse sistema de perceber-planejar-agir, ignora fenômenos importantes na tomada de decisão, como ações reflexas, e a velocidade com qual tais ações são expressadas em agentes cognitivos reais, mostrando, mais uma vez, que o método simbólico computacional não tem embasamento com a realidade biológica e física do fenômeno de cognição.

Herbert Dreyfus (1992) foi um dos mais proeminentes críticos da abordagem representacional simbólica no campo de pesquisa em IA. Com base na filosofia hermenêutico-existencialista proposta por Martin Heidegger, Dreyfus indicou em seus trabalhos que a manipulação de símbolos e de representações não são suficientes para gerar o tipo de existência não representacional de um ser-no-mundo (*Dasein*). No fundo deste impasse persiste uma crítica ao pensamento cartesiano materialista e ao dualismo sujeito-objeto: o cartesianismo materialista que tenta, sem sucesso, replicar o mundo inteiro "dentro da mente", está fadado a falhar, de acordo com Dreyfus, pois é impossível conter o mundo dentro da mente, pelo simples fato de que o mundo é infinitamente complexo e nós somos criaturas finitas (DREYFUS, 2007). Assim, um sistema autocontido, rígido, não é capaz de duplicar o tipo de agente cognitivo que desejamos. Talvez isso nos indique que representações e experiência devem operar em conjunto para que a primeira possua significado.

## Conexionismo e aprendizagem de valores

Podemos ver que muitos dos problemas citados anteriormente vêm da dificuldade dos programadores em expressarem o significado do que é proposto pela linguagem (erros de especificação) e como isso deve se modificar quando o contexto do ambiente evolui (erros de robustez). Seja pelo modelo cognitivo representacional, utilizando regras de comportamento (se-então), ou o modelo conexionista, utilizando redes neurais artificiais com funções de recompensa, ainda chegamos no mesmo impasse. Como expressar nossos objetivos e alinhar os valores de agentes inteligentes artificiais com os nossos?

A abordagem conexionista encontra diversas dificuldades nessa tarefa, sendo exploradas mais detalhadamente a seguir. Comumente redes neurais artificiais são treinadas de maneira supervisionada, utilizando dados de treinamento rotulados, entretanto este método pode não ser

o mais seguro para a aprendizagem de valores. Dreyfus e Dreyfus (1992), citam um exemplo onde um sistema de aprendizado de máquina treinado para classificar, ou não, veículos militares terrestres escondidos entre as árvores. O classificador durante o treinamento foi capaz de identificar, com grande precisão, os veículos desejados, porém, o sistema teve um desempenho pífio com imagens fora do grupo de treinamento. Descobriu-se posteriormente que o conjunto de fotos contendo veículos foram tiradas em um dia ensolarado, enquanto as imagens sem os veículos foram feitas em um dia nublado. O que o classificador realmente estava identificando era o brilho das imagens. Da mesma forma, a aprendizagem de valores por indução (treinamento supervisionado), é suscetível a esta falha (SOARES, 2016).

Por esse motivo espera-se que agentes inteligentes artificiais possuam uma propriedade chamada de corrigibilidade (*corrigibility*). É necessário que tais sistemas possam ter sua função de recompensa ou hierarquia de valores ajustada, caso algo indesejado ocorra. Contudo, também é necessário que os mesmos agentes não possam influenciar seu próprio quadro de aprendizado ou função de recompensa, muito menos impeçam que esta seja modificada. Atualmente não existem soluções para esse problema (SOARES; FALLENSTEIN, 2015).

Além disso, ambos os métodos de treinamento supervisionado, que usa dados rotulados e aprendizagem de reforço, que usa funções de recompensa como um *proxy* de resultados desejáveis, são extremamente vulneráveis na identificação de ambiguidades (SOARES, 2016), como atestam os problemas do "aprendiz de feiticeiro" (*Sorcerer's Apprentice*) e as situações em que o sistema, pela divergência nos ambientes de teste e os novos ambientes, tem a oportunidade de *hackear* sua recompensa (BOSTROM, 2014). O cenário de *hackeamento* de recompensa, ou "*wireheading*", é comparado erroneamente com humanos estimulando seu próprio prazer (por exemplo, uso de drogas). O apetite humano é saciável; um agente artificial com poder de maximizar sua própria recompensa não irá parar o seu comportamento "compulsivo". Inclusive, irá buscar formas e meios de perpetuar seu comportamento autocompensador livre de interferências (OMOHUNDRO, 2009).

A função de utilidade, ou função de recompensa como também é descrita na literatura, pode ser explicada pelo teorema da utilidade Von Neumann-Morgenstern (VON NEUMANN; MORGENSTERN, 1953). O teorema configura funções de utilidade através da ordenação de preferências: A é preferível a B, ou B é preferível a A, ou ambos possuem o mesmo valor de preferência. Uma função de utilidade permite que, dada o estado do agente e o estado do mundo em geral, se gere uma decisão do agente entre duas ou mais opções. O conceito de função de utilidade é uma formalização matemática para a noção de valores humanos, sendo amplamente utilizado em economia e teoria da decisão. Contudo, um dos problemas mais conhecidos desse modelo é o fato empírico que humanos violam os axiomas da teoria da utilidade, não tendo funções de utilidade consistentes (TVERSKY; KAHNEMAN, 1981).

Uma alternativa seria a modelação da intenção dos operadores utilizando aprendizagem por reforço inverso (NG; RUSSELL, 2000): onde um agente tenta identificar e maximizar a função de recompensa de algum outro agente no ambiente (geralmente um operador humano). Porém, preferências humanas não podem ser necessariamente capturadas apenas por observações, assim sistemas de aprendizagem por reforço inverso demonstram o problema de aprender "erros" ou "vícios" de comportamento humano como soluções válidas.

Avanços recentes na área, como o modelo de treinamento *Cooperative Inverse Reinforcement Learning* (CIRL) solucionariam esse problema: ao invés de estimar e adotar a função de recompensa do ser humano como sua própria, o sistema tenta resolver um *Partially observable Markov decision process* (POMDP), levando a um comportamento de aprendizagem cooperativa, no qual o sistema ou agente tenta maximizar a função de recompensa do operador, porém sem saber qual esta é (HADFIELD-MENELL, 2016). Contudo, essa abordagem gera problemas de interpretação, tais como a identificação de ambiguidade e os problemas de coordenação entre os agentes envolvidos no POMDP.

Ademais, situações onde humanos fazem parte do sistema de recompensa de uma IA, também chamados de *human-in-the-loop*, não são considerados seguros, pois existe forte evidência para se crer que agentes inteligentes artificiais seriam inclinados a manipular a parte humana de seu mecanismo de recompensa, se isso significasse uma aumento de recompensa (HIBBARD, 2012; BOSTROM, 2014).

Em geral, nossos métodos atuais de treinamento para o modelo cognitivo conexionista não são apropriadas para uma IA ou IAG (inteligência artificial geral) que opere no mundo real. Cenários possíveis de autoaperfeiçoamento, ou até mesmo uma "explosão de inteligência", como explicado pela Tese da Convergência Instrumental (BOSTROM, 2012), podem gerar consequências calamitosas para a humanidade (YUDKOWSKY, 2008). O objetivo final destes agentes é maximizar a recompensa, sendo nossos valores e metas apenas instrumentais para seu objetivo terminal. Tais agentes podem aprender que objetivos humanos são instrumentalmente úteis para altas recompensas, porém substituíveis, especialmente se a inteligência destes agentes for superior à nossa (DEWEY, 2011).

Seja pela representatividade simbólica, seja pelo treinamento conexionista, até o presente momento objetivos de valor não conseguem ser expressos de maneira segura, e dada a importância do alinhamento de valores humanos com IA, novos métodos devem ser investigados. Propomos, neste artigo, que o modelo cognitivo dinâmico oferece uma nova forma de pensar sobre o problema de alinhamento de valores. Na seção seguinte, discutiremos algumas das características do modelo teórico dinâmico de cognição.

### Modelo Teórico Cognitivo Dinâmico

Pode-se dizer que muitos modelos teóricos começam como metáforas ou analogias, tornando-se mais tarde teorias que podem ser implementadas em modelos e subsequentemente simuladas. As estruturas conceituais que formamos através desse processo podem ter grande impacto na forma que conduzimos nossos estudos, na maneira que abordamos o problema, a linguagem que descrevemos os fenômenos, e na maneira como formulamos uma pergunta, e interpretamos uma resposta. A teoria dos sistemas dinâmicos nos convida a pensar sobre o fenômeno da cognição e da experiência humana de uma forma progressista, tal como foi proposta por Van Gelder (1998, p. 4), cuja Hipótese Dinâmica postula: "Sistemas cognitivos naturais são certos tipos de sistemas dinâmicos, e são melhor compreendidos a partir da perspectiva dinâmica." Sistemas dinâmicos, nesse sentido, são sistemas em que, na medida que evoluem no tempo, suas variáveis estão continuamente e simultaneamente determinando a evolução uma das outras, em outras palavras, são sistemas regidos por equações diferenciais não lineares (GELDER; PORT, 1998, p. 6) Com esta afirmação, o dinamista coloca o agente em uma situação de acoplamento com o ambiente, tornando cérebro-corpo-ambiente um sistema dinâmico autônomo cognitivo onde não faz mais sentido falar de cognição ou experiência sem reconhecer os três aspectos desta tríade (GELDER; PORT, 1998, p. 23).

O modelo teórico que apresentamos nesse artigo é o SED (*Situated Embodied Dynamics*), proposto por Beer (2000), que enfatiza como a experiência cognitiva surge da interação dinâmica cérebro-corpo-ambiente. Em primeiro lugar, SED leva em conta a situação como sendo fundamental para a cognição, colocando a ação concreta, ou seja, literalmente, o agir no mundo, como algo mais fundamental do que a as descrições abstratas dessa ação. Assim, o trabalho final do agente inteligente é agir, ação esta que ocorre em um ambiente, o qual faz parte central do comportamento, pois é o que dá sentido e contexto à ação. E a interação do agente com o ambiente é mútua, não sendo o ambiente apenas uma fonte de problemas a serem resolvidos, mas um parceiro com o qual o agente está envolvido de momento a momento (FRANKISH; RAMSEY, 2014).

A atividade situada tem as suas origens filosóficas no trabalho fenomenológico de Heidegger (2012), que Dreyfus (1992) aplicou ao campo de IA, no qual se supõe que o agente heideggeriano não pode ser separado do ambiente ou do seu contexto interpretativo. A Psicologia Ecológica de Gibson (1979) também é um precursor da atividade situada, com sua noção de *affordances*: Gibson resalta a relação ambiente-organismo no fenômeno de percepção como uma via de mão dupla, onde se percebe para agir, e se age para perceber. A ideia de cognição situada pode ser ampliada para teorias como "mente estendida" (CLARK; CHALMERS, 1998), também conhecida como HEC (*hypothesis of extended cognition*) (ROCKWELL, 2010), que nos convida a pensar de uma maneira diferente em respeito ao pensamento cartesiano que coloca a mente aprisionada dentro do cérebro.

Ora, explicamos a gravidade como a relação entre os campos gravitacionais; o eletromagnetismo por campos eletromagnéticos; a posição de partículas subatômicas é expressa através de ondas probabilísticas usando a equação de Schrödinger, o comprimento de onda De Broglie e o princípio da incerteza de Heisenberg. Assim, parece provável que uma teoria sofisticada que explique a consciência e experiência do ser-no-mundo envolva algum tipo de teoria que faça referência à flutuação dinâmica de campos.

Em segundo lugar, na abordagem SED, a corporificação diz que a forma física e seus aspectos funcionais e biomecânicos são aspectos essenciais para o comportamento, como também sua biologia e fisiologia, no caso de agentes artificias, mecânica, *hardware* e *software*. Todos esses fatores criam a realização conceitual pela qual criamos nossas experiências e representações (FRANKISH; RAMSEY, 2014).

O pensamento da corporificação tem origem na fenomenologia trabalhada por Merleau-Ponty (1962), que foi, de resto, um dos precursores da noção de *affordance* de Gibson (1979), colocando o envolvimento corporal como crucial para a maneira como percebemos e agimos com o ambiente. Também sendo a estrutura biológica que suporta a cognição vital para o fenômeno cognitivo, devemos pensar sobre as implicações ou possibilidades desse fenômeno ser duplicado por componentes eletrônicos, e quais conceitos e abstrações tal formação poderia gerar, dado a importância da experiência corporal na criação de conceitos abstratos (LAKOFF; JOHNSON, 1999). Assim, fica patente o papel da linguagem, das metáforas e das representações mentais na própria formulação de conceitos utilizados em teorias científicas, a despeito de todo comprometimento ontológico com um certo realismo científico. Com efeito, o termo "epistemologia naturalizada", forjado por W. V. Quine em seu ensaio seminal de 1969, "Epistemology Naturalized", seguia várias das premissas epistêmicas do ceticismo de Hume, que, como assinalamos acima, solapa todo fundacionismo de inspiração platônica, incluindo o dualismo do racionalismo cartesiano, em sua pretensão de justificar um conhecimento absolutamente seguro da verdade do mundo exterior. De acordo com Quine (1969, p. 75),

> [...] era triste para epistemólogos, Hume e outros, ter de concordar com a impossibilidade de derivar estritamente a ciência do mundo externo de evidências sensoriais [*strictly deriving the science of the external world from sensory evidence*]. Dois princípios fundamentais do empirismo permaneceram inatacáveis, no entanto, e assim permanecem até hoje. Um é que qualquer evidência que exista para a ciência é evidência sensorial [*sensory evidence*]. O outro ...é que toda inculcação de significados de palavras deve repousar em última análise em evidências sensoriais.

Assim como em Quine, o empirismo de inspiração humeana que nos interessa, desde Dreyfus, Rorty, Prinz e o neopragmatismo, é intersubjetivo, falsificacionista e, interessantemente, externalista, isto é, uma forma de pragmatismo social, linguística e historicamente co-constitutivo de sujeito observador e de mundo objetivo a ser conhecido, experienciado, vivenciado, vivido. O problema do conhecimento, assim como o de dar razões para a ação moral, permanece o grande problema humano segundo a formulação humeana: nas palavras de Quine (1969, p. 72), "o problema humeano é o problema humano" (*the Humean predicament is the human predicament*), de forma que nem a indução (como a que tem sido adotada por modelos de equilíbrio reflexivo em metaética e filosofia da ciência) pode solucionar as falácias naturalistas que decorrem da guilhotina, justamente quando buscamos evitar uma dedução ou *mathesis universalis* da normatividade. O externalismo dos naturalistas, na esteira de Hume e Quine, se oporia aqui ao internalismo dos racionalistas e de Kant, segundo o qual a justificativa epistêmica para a cognição e para a ação moral encontra-se na consciência (*cogito*) ou em uma estrutura de subjetividade transcendental. Embora não possamos desenvolver aqui o problema internalista-externalista, cremos que o debate entre racionalismo e empirismo que o precedeu, nos autoriza a asserir, como Quine sugeriu, que o grande erro de Hume teria sido o de reduzir juízos analíticos a juízos a priori, universais necessários, em contraposição a juízos sintéticos, redutíveis por sua vez a juízos a posteriori, particulares contingentes, sem resolver o problema da indução, mas permitindo, ao contrário, o seu retorno pela porta dos fundos, como mostraria Popper, pelo autoengano de quem pretende justificar a ação moral com uma argumentação transcendental ou normativista. A nossa intuição programática sobre a ética da IA é, portanto, que nem o naturalismo parece lograr reduzir o alinhamento a um programa utilitarista nem os modelos deontológicos, normativistas e seus argumentos transcendentais parecem satisfatórios para evitar a suspeição antropomórfica.

A neuroetologia computacional é uma área distinta da neurociência computacional, pois envolve a criação de modelos conjuntos de circuitos neurais, biomecânicos e nichos ecológicos como partes relevantes de um agente cognitivo (CHIEL; BEER, 1997). Trabalhos no campo da robótica autônoma enfatizam que o comportamento inteligente é uma propriedade emergente de um agente incorporado em um ambiente com o qual deve interagir continuamente. Dessa forma, a visão computacional simbólica, que coloca o cérebro como a fonte de comandos que são emitidos para o corpo, pode estar incompleta. É possível que exista uma cognição ou "mente" do corpo (ou sistema mecânico), regida pelas próprias leis da física. Isso coloca o sistema nervoso não em uma posição de emitir comandos, mas sugestões, reconciliadas com o contexto biomecânico e ecológico (RAIBERT; HODGINS, 1993). Existe a possibilidade de que uma IA que possua compreensão de conceitos humanos exigiria um *design* muito próximo do de um ser humano (SOTALA; YAMPOLSKIY, 2013).

Por último, para compreendermos a abordagem SED devemos analisar a dinâmica pressuposta. Nos referimos à dinâmica como uma teoria matemática que descreve sistemas que mudam ao longo do tempo de uma maneira sistemática. O quadro dinâmico também nos fornece um filtro diferente para observar o fenômeno em questão (FRANKISH; RAMSEY, 2014). Um sistema dinâmico é uma abstração matemática composta de um espaço de estados *S*, um conjunto de tempo ordenado *T* e um operador de evolução que transforma um estado para outro ao longo de *T*. *S* pode ser numérico ou simbólico, contínuo, discreto ou híbrido, de qualquer topologia ou dimensão. *T* é tipicamente expressado pelo conjunto de números inteiros ou reais, e a evolução do operador pode ser determinística ou estocástica (KUZNETSOV, 2004).

Sistemas dinâmicos se configuram, decerto, como um corpo da matemática, e não como uma teoria científica do mundo natural. Os exemplos mais comuns de sistemas dinâmicos são conjuntos de equações diferenciais parciais, utilizadas para descrever fenômenos como o movimento da água, comportamento de campos eletromagnéticos, a posição de partículas subatômicas entre outros fenômenos naturais. Assim, a perspectiva dinâmica traz consigo um conjunto de conceitos e filtros que influenciam a forma que pensamos sobre o fenômeno estudado; quando se aproxima de algum sistema a partir da perspectiva dinâmica, busca-se identificar um conjunto de variáveis de estado cuja evolução possa explicar o comportamento observado, as leis dinâmicas pelas quais os valores dessas variáveis evoluem no tempo, a estrutura dimensional de sua evolução, possíveis estados e parâmetros dominantes (BEER, 2000).

Finalmente, a hipótese da estrutura situada, incorporada e dinâmica postula que cérebros, corpos e ambientes são sistemas dinâmicos, regidos por leis dinâmicas, e as dinâmicas dessa tríade estão acopladas, sendo o estudo do comportamento do sistema dinâmico completo, cérebro-corpo-ambiente, o objeto por excelência da estrutura SED (BEER, 2000). A conclusão mais crucial a ser tirada deste modelo é: o comportamento é uma propriedade do todo o sistema cérebro-corpo-ambiente e não pode, portanto, ser adequadamente atribuído a qualquer subsistema isoladamente dos outros. Dessa forma, propomos que tal abordagem, SED, poder ser um modelo interessante de aprendizagem de valores a se implementar em sistemas de IA.

### Discussão à guisa de conclusão

Como esta abordagem dinâmica pode ser útil para o problema de aprendizagem de valores? Essa tem sido a questão diretriz de nosso programa de pesquisa. Vimos nesse estudo o eminente avanço das tecnologias de IA, e a importância de que tais avanços sejam realizados de maneira segura, pois não podemos antropomorfizar IA, e esperar que agentes inteligentes artificiais possuam os mesmos objetivos terminais (valores) que nós. Portanto, a aprendizagem de valores se torna uma área de crucial importância no campo.

As limitações presentes no método simbólico representacional e no modelo conexionista podem estar nos indicando que uma abordagem diferente para os problemas de comportamento de agentes inteligentes deve ser cogitada. O dinamismo decerto aborda o problema de uma maneira diferente, e desvela novos aspectos que tanto o modelo simbólico quanto o conexionista deixam de lado.

Como devemos entender a natureza e o papel desse estado interno dentro de um agente dinâmico? A interpretação computacional tradicional de tais estados seria como representações internas. Infelizmente, apesar do papel fundamental que a noção de representação desempenha em abordagens computacionais, há muito pouco acordo sobre o que sua real função no controle e manutenção do comportamento. Devemos lembrar também que simbolismo, conexionismo e dinamismo são estruturas teóricas, não teorias científicas do mundo natural, ou seja, não podem ser provadas ou refutadas. Enquanto o simbolismo enfatiza a manipulação de representações internas, o conexionismo ressalta a arquitetura da rede e o protocolo de treinamento. Já a estrutura SED destaca o espaço de trajetória e as influências determinantes ao sistema cérebro-corpo-ambiente. É possível que uma abordagem dinâmica para o problema de aprendizagem de valores nos ajude a elucidar alguns dos problemas em aprendizagem de valores. Contudo, como dito anteriormente, não nos colocamos em uma posição de antirrepresentacionalismo. Ao contrário, é provável que uma teoria completa de cognição utilize as três estruturas teóricas. Sugerimos que em certos casos, como em comportamento orientado por objetivos, o funcionamento interno de um agente dinâmico não pode ser interpretado como representacional, a não ser que refinemos o que uma representação realmente pode ser ou significar.

Gärdenfors (2000), propõe uma teoria geral de representação, onde conceitos como valores são representados como formas geométricas dentro de um espaço multidimensional. Diversos estudos de modelagem cerebral tentam entender como o cérebro cria e manipula informações (KRIEGESKORTE; KIEVIT, 2013), e achados recentes utilizando simulações de grupos corticais analisados por topologia algébrica mostram que o cérebro parece se organizar de forma ordenada e geométrica quando analisamos sua estrutura como um objeto multidimensional (REIMANN *et al.*, 2017). É possível que estruturas similares, correspondentes ao conceito de valor, sejam encontradas no campo hiperdimensional que compõem o agente cognitivo.

A abordagem dinâmica se difere dos modelos cognitivos simbólico e conexionista pois coloca a biomecânica e ecologia com a mesma relevância que a atividade neural no fenômeno de cognição. Talvez, as dificuldades que temos encontrado na aprendizagem de valores e outros problemas no campo de IA se dê pelo fato de estarmos ignorando dois fatores cruciais do fenômeno. As implicações da hipótese dinâmica não só trazem uma nova forma de pensar como novos problemas para o campo de pesquisa em IA nutrindo, assim, novas ideias em áreas como neurofilosofia, neurociência, metaética, neuroetologia computacional e o próprio campo interdisciplinar da ciência cognitiva. Em conclusão, é necessária uma melhora na educação em conceitos de sistemas dinâmicos, com a promessa de que tais métodos possam ser úteis para o problema de alinhamento de valores em IA e para comunidade de ciência cognitiva em geral.

## Referências

**Nicholas Kluge Corrêa**

Mestre em Engenharia Elétrica pela Pontifícia Universidade Católica do Rio Grande do Sul (Escola Politécnica, PUCRS, Porto Alegre, RS, Brasil) e doutorando em Filosofia (PUCRS) Porto Alegre, RS, Brasil. Bolsista CAPES/PROEX.

**Nythamar de Oliveira**

Ph.D. in Philosophy (State University of New York). Professor titular da Pontifícia Universidade Católica do Rio Grande do Sul (PUCRS), Porto Alegre, RS, Brasil).

**Endereço para correspondência**

**Nicholas Kluge Corrêa/ Nythamar de Oliveira**.

Pontifícia Universidade Católica do Rio Grande do Sul Av. Ipiranga, 6681, Prédio 8, 4.º andar, Partenon, 90619900, Porto Alegre, RS, Brasil